\documentclass{article}
\usepackage{arxiv}
\usepackage[utf8]{inputenc} 
\usepackage[T1]{fontenc}    
\usepackage{hyperref}       
\usepackage{url}            
\usepackage{booktabs}       
\usepackage{amsfonts}       
\usepackage{nicefrac}       
\usepackage{microtype}      
\usepackage{lipsum}		
\usepackage{graphicx}
\usepackage{natbib}
\usepackage{doi}
\usepackage{booktabs}
\usepackage{tcolorbox}
\usepackage{geometry}
\usepackage{enumitem}
\usepackage{float}

\usepackage{algpseudocode}
\title{ \textsc{\Large Optimising Supertrend Parameters using\\ Bayesian Optimisation for Maximising profit and other metrics}}

\author{ \href{https://sites.google.com/view/shafiq-abdulrahman-iitm/home}{\hspace{1mm}Abdul Rahman}{} \\
	Department of Mathematics\\
	Indian Institute of Technology,Madras\\
	IITM,Chennai,India \\
	\texttt{ma22c001@smail.iitm.ac.in} \\
}

\renewcommand{\headeright}{}
\renewcommand{\undertitle}{}
\renewcommand{\shorttitle}{ }

\hypersetup{
pdftitle={A template for the arxiv style},
pdfsubject={q-bio.NC, q-bio.QM},
pdfauthor={David S.~Hippocampus, Elias D.~Striatum},
pdfkeywords={First keyword, Second keyword, More},
}

\begin{document}
\maketitle

\begin{abstract}
	This thesis  investigates the potential of Bayesian optimization (BO) to optimize the atr multiplier and atr period -the parameters of the Supertrend indicator for maximizing trading profits across diverse stock datasets. By employing BO, the thesis  aims to automate the identification of optimal parameter settings, leading to a more data-driven and potentially more profitable trading strategy compared to relying on manually chosen parameters. The effectiveness of the BO-optimized Supertrend strategy will be evaluated through backtesting on a variety of stock datasets.
\end{abstract}


\section{Introduction}

\underline{Unveiling Profit Potential through Supertrend Parameter Optimization with Bayesian Techniques}

The financial markets, with their intricate dynamics and perpetual evolution, have long captivated scholars and practitioners alike. While the quest for a foolproof prediction method remains an elusive ideal, technical analysis offers a valuable toolkit for navigating market trends and identifying potential entry and exit points for investment strategies. Within this realm, the Supertrend indicator has emerged as a popular choice for its simplicity and effectiveness in capturing prevailing market directions.

The Supertrend indicator, built upon the foundation of the Average True Range (ATR), provides visual cues regarding the dominant market trend. It constructs a dynamic line, incorporating a user-defined multiplier, that adapts to market volatility. When the price action consistently pushes above the Supertrend line, it suggests a potential uptrend, prompting the adoption of long positions. Conversely, a price consistently trading below the line signals a possible downtrend, favoring short positions.

Despite its apparent utility, the efficacy of the Supertrend indicator hinges on two crucial parameters: the atr\_multiplier and the atr\_period. The atr\_multiplier dictates the sensitivity of the Supertrend line to price fluctuations. A higher multiplier leads to a steeper line, potentially generating fewer but potentially larger signals, while a lower multiplier creates a smoother line with potentially more frequent but smaller signals. The atr\_period, on the other hand, influences the responsiveness of the Supertrend line to recent price movements. A shorter period captures recent trends quicker but might be more susceptible to volatility, while a longer period provides a more smoothed-out trend but might lag behind rapid price changes.

Striking an optimal balance between these parameters is critical for maximizing a strategy's effectiveness. However, manually testing every possible combination is not only time-consuming but also inefficient, especially as the number of parameters increases. This is where Bayesian optimization (BO) enters the scene as a powerful and sophisticated optimization technique.

Bayesian optimization is a data-driven approach to efficiently exploring a vast parameter space to identify the settings that yield the best results. Unlike traditional methods such as grid search or random search, BO leverages past evaluations to prioritize promising areas of the search space.(\cite{article})
This iterative process allows BO to converge on the optimal parameter combination significantly faster, making it an ideal choice for fine-tuning trading strategies.(\cite{Au2023})

This thesis  delves into the potential of exploiting Bayesian optimization to optimize the atr\_multiplier and atr\_period of the Supertrend indicator for maximizing trading profits across a diverse set of stock datasets. Also, aiming to contribute to the discourse surrounding algorithmic trading by exploring:

\begin{itemize}
    \item The theoretical underpinnings of the Supertrend indicator and its role within the technical analysis framework.
    \item The critical role of hyperparameter tuning in trading strategies and the inherent limitations of traditional optimization methods.
    \item The fundamental principles of Bayesian optimization and its advantages in efficiently navigating large parameter spaces.
    \item The application of BO to optimize Supertrend parameters for various stock datasets, encompassing both domestic and international markets.
    \item A comprehensive evaluation of the optimized strategies' performance using established metrics like profit factor and maximum drawdown.
\end{itemize}

By harnessing the combined strengths of the Supertrend indicator and Bayesian optimization, this thesis aspires to unveil the potential for enhanced profitability within the algorithmic trading landscape. Further explored whether BO can unlock the hidden potential within the Supertrend framework, potentially leading to strategies that consistently outperform the baseline. However, it is crucial to acknowledge that the pursuit of maximum profits inherently involves risk management considerations. Also addressesing these trade-offs and emphasizing the importance of backtesting and risk mitigation strategies before deploying any optimized strategy in a real-world trading environment.(\citep{hadash2018estimate}). This formal introduction sets the stage for our exploration. In the subsequent sections, the thesis delves deeper into the relevant scholarly literature surrounding the Supertrend indicator and Bayesian optimization, establishes the problem statement and data considerations, and outlines the methodologies employed to optimize and evaluate the Supertrend-based strategies.

\section{Literature review}
\subsection{Supertrend Indicator: A Trend Following Powerhouse}

The Supertrend indicator has carved a niche in the realm of technical analysis as a straightforward yet effective tool for identifying trends. Built upon the foundation of the Average True Range (ATR), it constructs a dynamic line that adapts to market volatility, offering visual cues regarding the prevailing market direction (\cite{sahoo2017performance}).

\subsubsection{Calculation}

The Supertrend calculation incorporates two user-defined parameters:

\begin{itemize}
    \item \textbf{atr\_multiplier :} This factor determines the sensitivity of the Supertrend line to price fluctuations. A higher multiplier leads to a steeper line, potentially generating fewer but potentially larger signals.This is a constant value that traders and investors employ to push the indicator to be more or less sensitive to price movements. Generally, whole numbers are used, but the multiplier can be made more specific to fit the trading strategy
    \item \textbf{atr\_period :} This parameter influences the responsiveness of the Supertrend line to recent price movements. A shorter period captures recent trends quicker but might be more susceptible to volatility, while a longer period provides a more smoothed-out trend but might lag behind rapid price changes.The ATR period is calculated based on the highest and lowest prices, as well as the closing price of the asset over a specified time frame.
\end{itemize}

The Supertrend line (ST) is calculated using the following formula (\cite{bhuiyan2016comparative}):
\begin{tcolorbox}

 \[  Supertrend=\frac{(High+Low)}{2} + (ATR\_Multiplier)
 \times ATR\_Period \]

\end{tcolorbox}
\begin{itemize}
\item \textbf{High and low:} These are the highest and lowest prices of the asset during a specified time frame.
\end{itemize}

\subsubsection{Advantages}

\begin{itemize}
    \item \textbf{Simplicity:} The Supertrend indicator boasts a straightforward calculation, making it readily interpretable for traders of all experience levels \citep{sahoo2017performance}.
    \item \textbf{Trend Identification:} It's dynamic nature allows for effective identification of both uptrends (price consistently above the Supertrend line) and downtrends (price consistently below the Supertrend line) \citep{bhuiyan2016comparative}.
    \item \textbf{Reduced False Signals:} Compared to some other trend-following indicators, the Supertrend may generate fewer false signals due to its incorporation of the ATR, which helps account for volatility .
\end{itemize}
    
\subsubsection{Limitations:}
\begin{itemize}
\item Lagging Indicator: The Supertrend indicator, like most technical indicators, is inherently backward-looking, relying on past price data to generate signals. This can lead to delayed entry and exit points, particularly during volatile market environments \citep{sahoo2017performance}
\item Whipsaws: During periods of increased market volatility, the Supertrend indicator may generate whipsaws, where the price oscillates above and below the Supertrend line, leading to unnecessary trades and potential losses.
\item Parameter Sensitivity: The Supertrend indicator's performance is heavily influenced by the chosen atr multiplier and atr period. Finding the optimal settings can be challenging and may require experimentation or optimization techniques \citep{bhuiyan2016comparative}.
\end{itemize}
\subsubsection{Hyperparameter Optimization:} 
\underline{Fine-Tuning for Enhanced Performance}

The realm of algorithmic trading hinges on the crucial concept of hyperparameter tuning. These parameters, distinct from the data itself, govern how a trading strategy operates. In the context of the Supertrend indicator, the atr multiplier and atr period are prime examples of hyperparameters. Identifying their optimal settings is paramount for maximizing a strategy's effectiveness \citep{osorio2020optimizing}.

Traditional optimization methods, such as grid search or random search, can be employed to explore the parameter space. However, these methods can become computationally expensive, especially as the number of hyperparameters increases.\citep{greenblatt2016bayesian}

Bayesian Optimization (BO) emerges as a powerful alternative. This data-driven approach efficiently navigates large parameter spaces by leveraging past evaluations to prioritize promising areas for exploration. This iterative process allows BO to converge on the optimal parameter combination significantly faster than traditional methods .\citep{greenblatt2016bayesian}

\subsection{Bayesian Optimization: A Data-Driven Path to Optimal Parameters}

In the realm of algorithmic optimization, Bayesian optimization (BO) emerges as a powerful and sophisticated technique. It transcends the limitations of traditional grid search or random search methods by leveraging a probabilistic approach to efficiently navigate vast parameter spaces. Unlike its brute-force counterparts, BO prioritizes promising areas of exploration based on past evaluations, leading to a faster convergence on the optimal parameter combination.

\textbf{Definition:}

Bayesian optimization can be defined as an iterative process for optimizing an unknown objective function by building a probabilistic model of its behavior. It utilizes a combination of:

\begin{itemize}
    \item Gaussian Process (GP): A flexible probabilistic model that captures the relationship between input parameters (hyperparameters) and the resulting objective function values.
    \item Acquisition Function: This function guides the search process by balancing exploration (discovering new areas) and exploitation (focusing on promising areas) based on the current understanding of the objective function. Common acquisition functions include Expected Improvement (EI) and Upper Confidence Bound (UCB).
\end{itemize}

\subsubsection{\textbf{Algorithm: }}

\begin{enumerate}
    \item Initialization: Choose an initial set of parameter combinations and evaluate the objective function for each.
    \item Model Building: Train a Gaussian Process model based on the collected data (hyperparameters and corresponding objective function values).
    \item Acquisition Function Optimization: Utilize the acquisition function to identify the next most promising parameter combination to evaluate.
    \item Evaluation: Evaluate the objective function for the chosen parameter combination.
    \item Iteration: Update the Gaussian Process model with the new data point and repeat steps 3-5 until a stopping criterion is met (e.g., reaching a maximum number of evaluations).
\end{enumerate}

\subsubsection{\textbf{Benefits for Finance and Parameter Optimization:}}
\begin{itemize}
    \item Efficiency: Compared to traditional methods, BO significantly reduces the number of evaluations needed to reach the optimal parameter settings, especially for problems with high-dimensional parameter spaces.
    \item Global Optimization: BO avoids getting stuck in local optima by balancing exploration and exploitation during the search process.
    \item Data-Driven Approach: BO leverages past evaluations to improve its understanding of the objective function and guide the search towards better solutions.
\end{itemize}

\subsubsection{\textbf{Importance in Algorithmic Trading:}}
\begin{itemize}
    \item Hyperparameter Tuning: BO excels at optimizing hyperparameters in algorithmic trading strategies, such as the \texttt{atr\_multiplier} and \texttt{atr\_period} of the Supertrend indicator. By finding the optimal settings, BO can potentially enhance the strategy's performance and profitability.
    \item Portfolio Allocation: BO can be applied to optimize portfolio allocation across various assets, balancing risk and return based on historical data and market conditions.
    \item Risk Management: BO can be used to optimize risk management parameters, such as stop-loss and take-profit levels, helping to mitigate potential losses.
\end{itemize}


\section{Problem Statement}
\begin{tcolorbox}
This research thesis investigates the potential of leveraging Bayesian optimization (BO) to optimize the performance of the Supertrend indicator for maximizing trading profits across a diverse set of stock datasets. Specifically, aim is to determine whether BO can identify the optimal \texttt{atr\_multiplier} and \texttt{atr\_period} parameters for the Supertrend indicator, leading to a more effective strategy for generating buy and sell signals.
\end{tcolorbox}
\subsection{Data Description}

The research will employ historical price data encompassing a mix of domestic and international stocks to evaluate the effectiveness of the BO-optimized Supertrend strategy. The chosen datasets include:

\begin{itemize}
    \item Nifty 50: This index represents the top 50 companies listed on the National Stock Exchange of India.
    \item Infosys Limited: A leading Indian multinational information technology company.
    \item Hindustan Unilever Limited: An Indian multinational consumer goods company.
    \item Microsoft Corporation: A multinational technology company headquartered in the United States.
    \item NVIDIA Corporation: An American multinational technology company that designs graphics processing units (GPUs) for the gaming and professional markets.
\end{itemize}

\subsection{Data Acquisition}

The historical price data for these stocks will be acquired from a reliable financial data source through an appropriate API or service like \textbf{Yahoo Finance.} The specific timeframe for the data will be determined based on data availability and the research objectives. Daily, weekly, or monthly data can be considered depending on the chosen trading strategy and desired level of granularity.

\subsection{Data Preprocessing}

Essential data cleaning and preprocessing steps will be undertaken before applying the Supertrend indicator and BO. These steps may include:

\begin{itemize}
    \item Missing Value Handling: Missing price data points will be addressed using appropriate techniques like forward fill, interpolation, or data removal if the missing data is significant.
    \item Normalization (Optional): Depending on the chosen BO library and objective function, the price data may need normalization (e.g., scaling to a range of 0-1) to ensure all features contribute equally during optimization.
\end{itemize}

\subsection{Trading Strategy based on Supertrend}

The trading strategy will utilize the Supertrend indicator to generate buy and sell signals:
\begin{tcolorbox}[colback=red!5!white,colframe=red!75!black,title=  Strategy ]
\begin{itemize}
    \item Buy Signal: A long position is initiated when the closing price of a stock crosses above the Supertrend line (Upper Band - UB). This suggests a potential uptrend.
    \item Sell Signal: A short position is entered (or a long position is exited) when the closing price crosses below the Supertrend line (Lower Band - LB). This indicates a potential downtrend.
\end{itemize}
\end{tcolorbox}
\subsection{Additional Considerations}

By implementing BO, this research aims to identify the optimal atr multiplier and atr period values for the Supertrend indicator within each dataset. This optimization process can potentially lead to a more robust and profitable trading strategy while acknowledging the importance of risk management principles.

\subsection{Bayesian Optimization for Supertrend Parameter Tuning}
\subsubsection{Functionality}
This section explores the application of Bayesian optimization (BO) for identifying optimal parameter settings for the \texttt{atr\_multiplier} and \texttt{atr\_period} of the Supertrend indicator. BO operates iteratively:

\begin{enumerate}
  \item \textbf{Search Space Definition:} The permissible range of values for both \texttt{atr\_multiplier} and \texttt{atr\_period} is defined, creating a well-defined search space for exploration.

  \item \textbf{Objective Function:} A crucial element is the objective function, which evaluates the performance of a Supertrend strategy based on chosen parameters. It takes historical price data and parameter values, generates buy/sell signals based on the Supertrend indicator, and calculates a performance metric (e.g., profit factor).

  \item \textbf{Initial Exploration \& Iterative Improvement:} BO begins by evaluating the objective function for a predefined number of initial parameter combinations. This initial exploration informs a Gaussian Process model, which captures the relationship between parameters and objective function values. As evaluations accumulate, the model is refined, guiding the exploration process.

  \item \textbf{Acquisition Function \& Stopping Criteria:} An acquisition function prioritizes the selection of the next parameter combination for evaluation, balancing exploration and exploitation. Common functions include Expected Improvement (EI) and Upper Confidence Bound (UCB). The process continues until a stopping criterion (e.g., maximum evaluations or desired performance) is met.
\end{enumerate}

\section{Methodology:}

\subsection{Implementing the BO-Optimized Supertrend Strategy}
This section delves into the methodological approach for applying the BO-optimized Supertrend strategy. Further, outlining the algorithmic steps and Pseudocode for key functionalities.

\definecolor{mycolor}{RGB}{255,215,0}

\begin{tcolorbox}[colback=mycolor!10,colframe=mycolor,title={Methodology: Complete Algorithm},fonttitle=\bfseries]

Define the target stock list, such as Nifty 50, Infosys, etc.\\

Obtain historical price data for each stock using a financial data API, such as Yahoo Finance.\\

Preprocess the data by handling missing values and performing normalization if needed.\\

\textbf{Supertrend Indicator Function}\\
Define a function \texttt{calculate\_supertrend(high, low, close, atr\_multiplier, atr\_period)}\\
Calculate the Average True Range (ATR) using the chosen atr\_period.\\
Initialize the Supertrend line with the first period's data.\\
For each remaining data point:\\
\quad Update the Supertrend line based on the closing price relative to the previous Supertrend and ATR values.\\

\textbf{Bayesian Optimization Function}\\
Define a function to optimize Supertrend parameters (data, objective function)\\
Utilize a BO library (e.g., sci-kit-optimize) to define the search space for atr multiplier and atr period.\\
Implement the chosen objective function (e.g., profit factor) to evaluate a Supertrend strategy with specific parameters on the provided data.\\
Run the BO algorithm to identify the combination of atr multiplier and atr period that maximizes the objective function.\\

\textbf {Strategy Backtesting}\\
Use the optimized atr multiplier and atr period for each stock.\\
Simulate the trading strategy based on Supertrend buy/sell signals.\\
Evaluate the backtesting performance using metrics like profit factor, drawdown, and Sharpe ratio.\\

\textbf{Evaluation and Refinement}\\
Analyze the backtesting results and consider incorporating additional filters or risk management techniques.\\
Refine the objective function or BO parameters if necessary.

\end{tcolorbox}


\section{Empirical analysis}
\subsection{Strategy execution and performance}
This section presents the empirical results obtained by applying the Supertrend strategy with default atr multiplier (3) and atr period (15) to the collected datasets for the five stocks: Nifty 50, Infosys, Hindustan Unilever, Microsoft, and Nvidia.
\begin{tcolorbox}
\begin{center} 

\href{https://colab.research.google.com/drive/18skAFcDahTUyxcHvBUKu9RHFmiyWX8kU?usp=sharing}{ \textcolor{blue}{https://colab.research.google.com/drive.supertrendstrategy}}

\end{center}   
\end{tcolorbox}

\subsection{Performance of Metric:}
The backtesting process generated key performance metrics for each stock dataset. These metrics include profit factor,  maximum drawdown, Overall Profit /Loss percentage, Minimum and Maximum balance when one invests 100 units, and total trades during the time frame.

\begin{table}[htbp]
  \centering
  \caption{Performance Metrics using default parameters}
  \begin{tabular}{|l|c|c|c|c|c|}
  \hline
    & \textbf{Nifty 50} & \textbf{Infosys} & \textbf{HUL} & \textbf{Microsoft} & \textbf{Nvidia} \\
    \hline
   Overall P/L \% & -1.16\% & 8.25\% & 1.37\% & 1.61\% & 6.64\% \\
    Overall P/L & -1.16 & 8.25 & 1.37 & 1.61 & 6.64 \\
    Min Balance & 93.36 & 100.00 & 100.00 & 98.25 & 98.91 \\
    Max Balance & 100.00 & 111.96 & 108.30 & 108.02 & 110.67 \\
    Max Drawdown & -6.64 & -4.69 & -6.93 & -6.62 & -8.73 \\
    Max Drawdown \% & -6.64\% & -4.19\% & -6.40\% & -6.13\% & -7.89\% \\
    Total Trades & 37 & 37 & 27 & 32 & 38 \\
    \textcolor{red}{Max Profit} & \textcolor{red}{0.00} & \textcolor{red}{11.96} & \textcolor{red}{8.30} & \textcolor{red}{8.02} & \textcolor{red}{10.67} \\
    \hline
  \end{tabular}
\end{table}

\newpage

\begin{center}
    Indian Stocks and their Strategy Performance 
\end{center}

\begin{figure}[htbp]
  \begin{minipage}[t]{0.3\linewidth}
    \includegraphics[width=\linewidth]{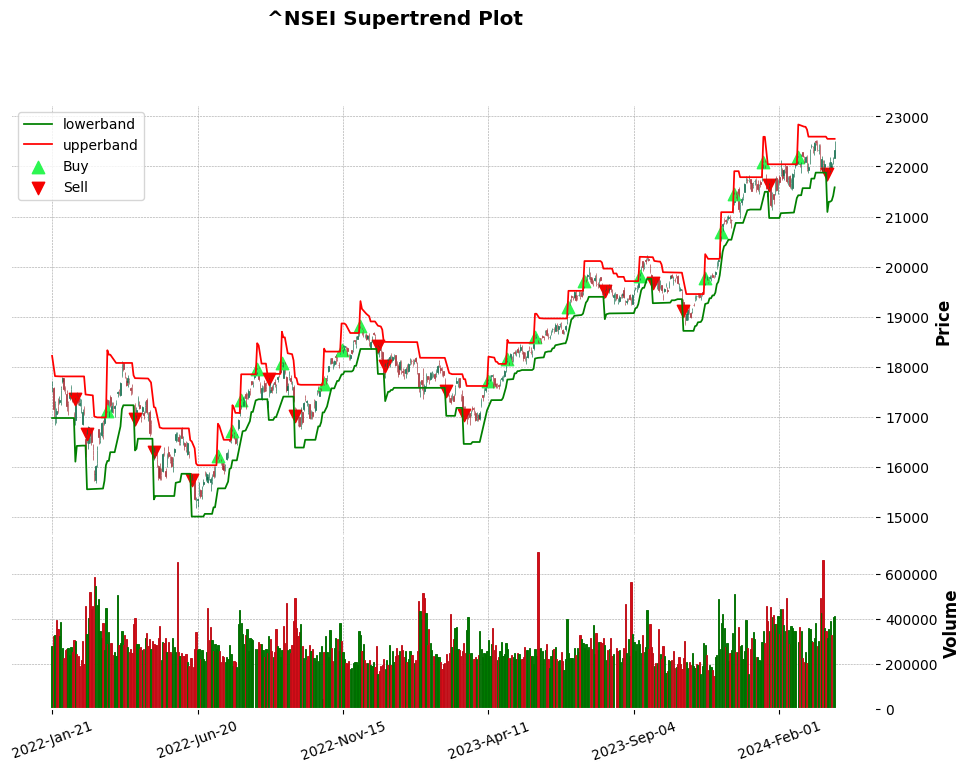}
    \caption{Nifty 50}
  \end{minipage}
  \hfill
  \begin{minipage}[t]{0.3\linewidth}
    \includegraphics[width=\linewidth]{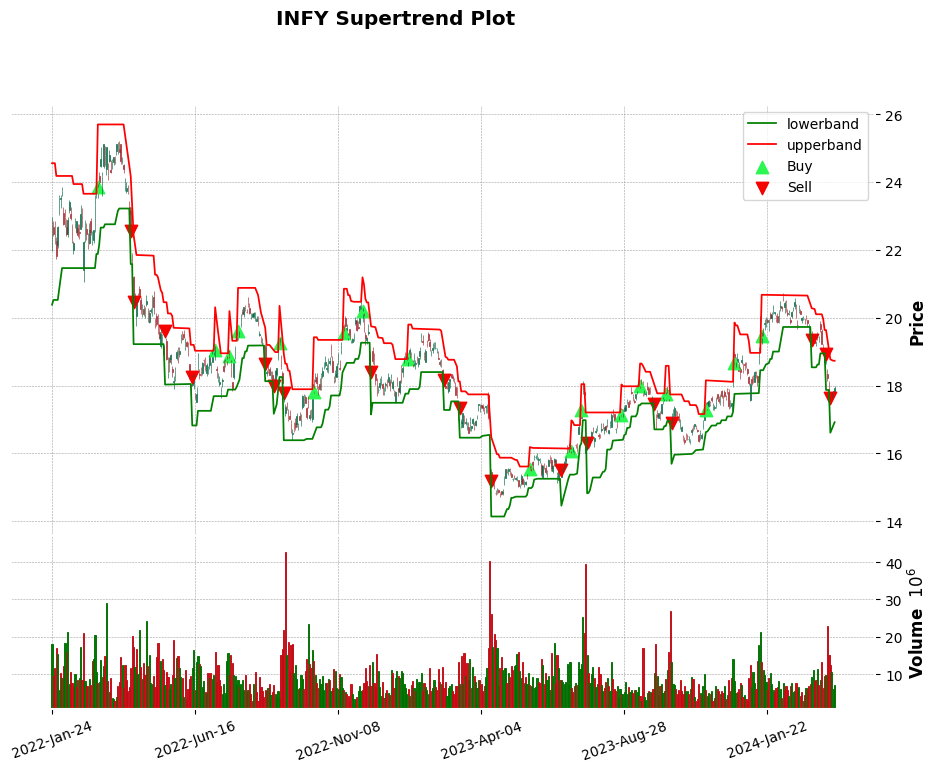}
    \caption{Infosys}
  \end{minipage}
  \hfill
  \begin{minipage}[t]{0.3\linewidth}
    \includegraphics[width=\linewidth]{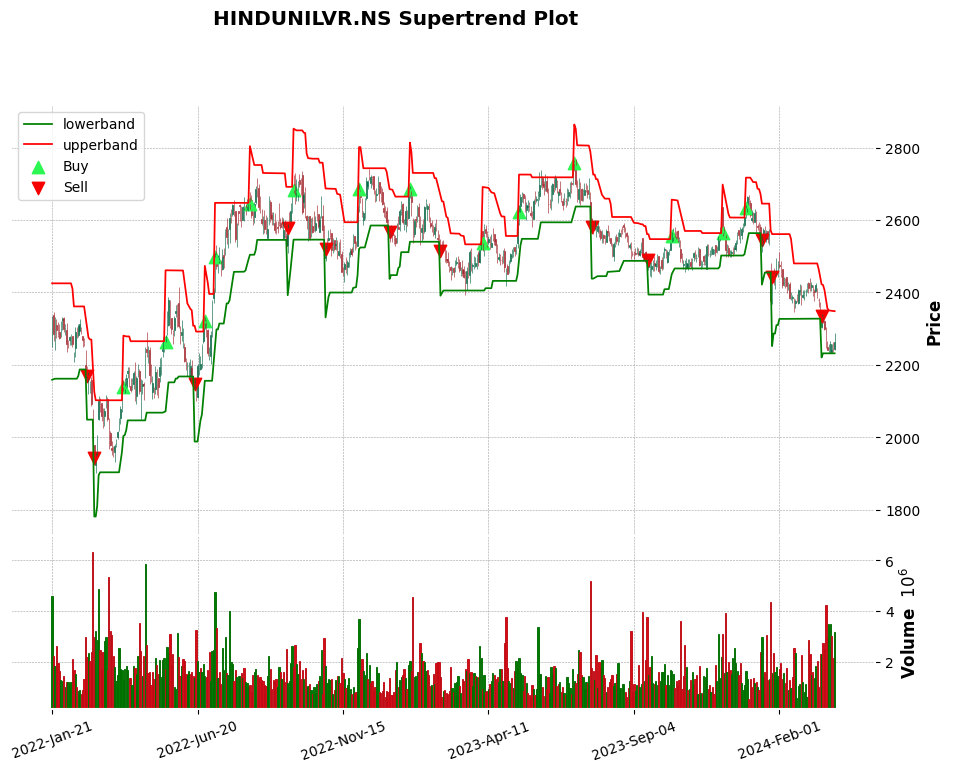}
    \caption{Hindustan Unilever Ltd}
  \end{minipage}
\end{figure}

\subsubsection{Explanation of Supertrend Strategy Performance Metrics with default parameters}
The above table(1) summarizes the performance of the Supertrend strategy applied to five assets: Nifty 50, Infosys, Hindustan Unilever (HUL), Microsoft, and Nvidia. Here's a breakdown of the key metrics and observations:

\begin{itemize}
    \item \textbf{Overall Profit or Loss (P/L):}
    \begin{itemize}
        \item \textbf{Percentage (\%):} The first row (-1.16\%) represents the overall loss of the strategy across all five assets during the backtesting period.
        \item \textbf{Value:} The second row (-1.16) shows the same loss expressed as a numerical value.
    \end{itemize}
    
    \item \textbf{Balance Metrics:}
    \begin{itemize}
        \item \textbf{Min Balance:} This indicates the lowest point reached by the cumulative return of each asset during backtesting. For example, Nifty 50 experienced a minimum balance of 93.36, suggesting a 6.64\% decline from its starting point.
        \item \textbf{Max Balance:} This represents the highest point reached by the cumulative return for each asset. Infosys achieved the highest balance (111.96), indicating an 11.96\% gain from its starting point.
    \end{itemize}
    
    \item \textbf{Drawdown Metrics:}
    \begin{itemize}
        \item \textbf{Max Drawdown:} This metric shows the peak decline in the cumulative return of each asset relative to its highest point. For example, Nvidia experienced a maximum drawdown of -8.73, signifying a decline of 8.73\% from its peak balance.
        \item \textbf{Max Drawdown (\%):} This expresses the maximum drawdown as a percentage, offering a relative comparison across assets. Nvidia again has the highest drawdown percentage (-7.89\%).
    \end{itemize}
    
    \item \textbf{Trading Activity:}
    \begin{itemize}
        \item \textbf{Total Trades:} This indicates the total number of trades executed by the strategy for each asset during the backtesting period. Nifty 50 and Infosys both had 37 trades, while Nvidia had the highest number of trades (38).
        \\
    \end{itemize}
\end{itemize}

\textbf{Important Points:}
\begin{itemize}
    \item Despite some individual assets generating profits (Infosys, Microsoft, Nvidia), the overall strategy resulted in a slight loss (-1.16\%).
    \item Infosys exhibited the strongest performance with a maximum balance of 111.96 (8.25\% gain) and a relatively low maximum drawdown (-4.69\%).
    \item Nvidia had the highest number of trades (38) but also experienced the most significant drawdown (-8.73 or -7.89\%).
\end{itemize}

\textbf{Further Analysis:}
\begin{itemize}
    \item Investigating the reasons behind the overall loss and the varying performance across assets would be valuable for potential strategy refinements.
    \item By delving deeper into these metrics and incorporating additional performance measures, investor can gain a richer understanding of the \textcolor{blue}{Supertrend strategy's effectiveness with default parameters and identify areas for potential improvement through Bayesian optimization.}
\end{itemize}

\begin{center}
    International Stocks and their strategy performance 
\end{center}
\begin{figure}[htbp]
\centering
  \begin{minipage}[t]{0.4\linewidth}
    \includegraphics[width=\linewidth]{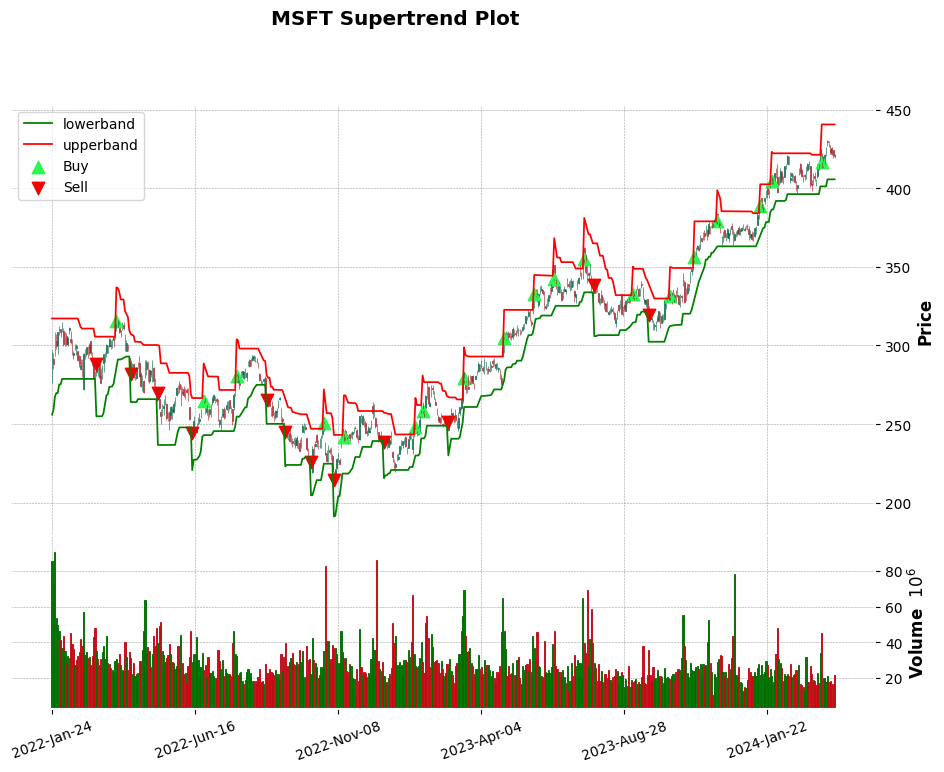}
    \caption{Microsoft Corporation}
  \end{minipage}
  \hfill
  \begin{minipage}[t]{0.4\linewidth}
    \includegraphics[width=\linewidth]{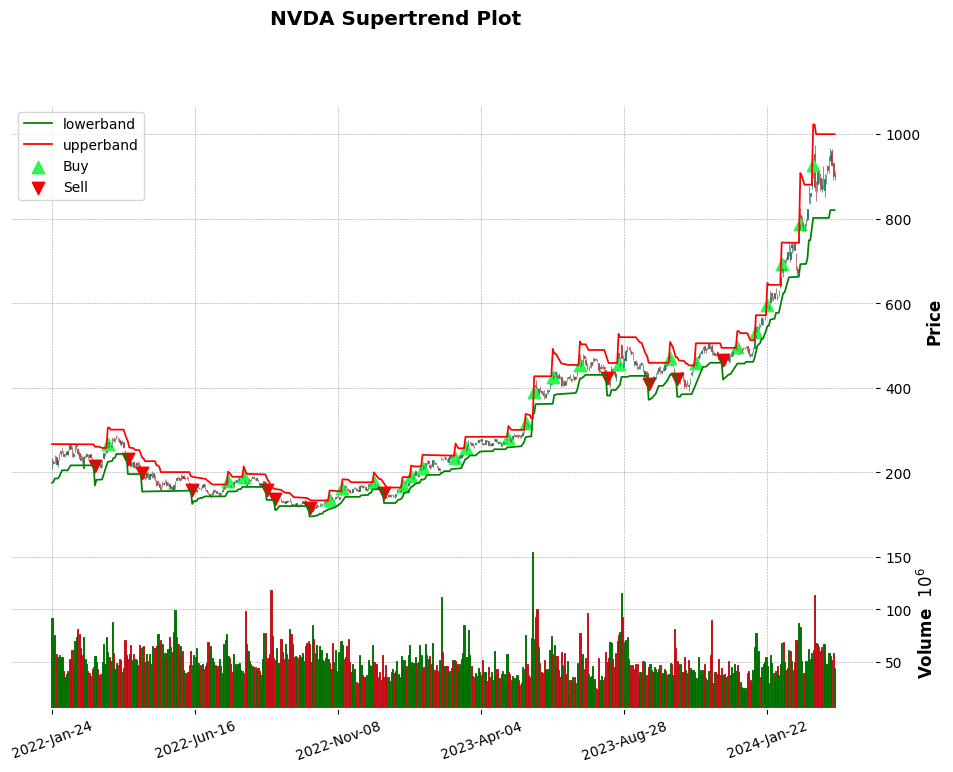}
    \caption{Nvidia}
  \end{minipage}
\end{figure}



\section{ Bayesian Optimization for Supertrend Parameter Tuning to optimise profit factor}

This section explores the application of Bayesian optimization (BO) to optimize the \texttt{atr\_multiplier} and \texttt{atr\_period} parameters of the Supertrend strategy for each individual stock dataset. The objective is to maximize the profitability of the strategy across diverse market conditions.

\subsection{ BO Framework}

\subsubsection{ Data Splitting}

The initial step involves splitting the historical price data for each stock into training and test sets using a train-test split. The training set (typically encompassing 70-80\% of the data) will serve two purposes:

\begin{itemize}[label=--]
    \item BO and Strategy Evaluation: The BO algorithm will utilize the training set to identify optimal parameter configurations for the Supertrend strategy. This involves evaluating the strategy's performance with different parameter combinations and selecting the one that maximizes the chosen objective function.
    \item Cross-Validation (Optional): To enhance the robustness of the BO process and mitigate overfitting, k-fold cross-validation can be implemented within the training set. This involves further dividing the training data into k folds, performing separate BO and evaluation iterations on each fold, and averaging the results. While not strictly necessary for BO itself, cross-validation can provide a more reliable estimate of the optimized parameters' generalizability.
\end{itemize}

The remaining 20-30\% of the data will constitute the test set. This unseen data serves for the final performance assessment of the Supertrend strategy after parameter optimization using BO.

\subsubsection{Objective Function}

The objective function plays a crucial role in BO, guiding the optimization process towards parameter combinations that maximize the desired performance metric. In this context, the objective function will focus on maximizing the profitability of the Supertrend strategy. Potential options include:

\begin{itemize}[label=--]
    \item Profit Factor: This metric is well-suited for this scenario as it considers both winning and losing trades, providing a comprehensive measure of profitability.
    \item Sharpe Ratio: If risk-adjusted returns are a priority, the Sharpe Ratio can be employed as the objective function. This metric incorporates both profitability and volatility, offering a more nuanced perspective on strategy performance.
\end{itemize}

The chosen objective function will be clearly defined and calculated for each parameter combination evaluated during the BO process.

\subsection{Pseudocode for Bayesian Optimization}



\definecolor{mycolor}{RGB}{255,215,0}

\begin{tcolorbox}[colback=mycolor!10,colframe=mycolor,title={Bayesian Optimization for Supertrend Parameter Tuning},fonttitle=\bfseries]

\textbf{function} objective(atr\_period, atr\_multiplier)\\
\quad \textbf{global} data, results\_dataframe(df)\\
\quad atr\_period $\gets$ int(atr\_period)\\
\quad atr\_multiplier $\gets$ int(atr\_multiplier)\\
\quad supertrend\_data $\gets$ supertrend(data.copy(), atr\_period=atr\_period, atr\_multiplier=atr\_multiplier)\\
\quad supertrend\_positions $\gets$ generate\_signals(supertrend\_data)\\
\quad strategy\_df $\gets$ strategy\_performance(supertrend\_positions, capital=100, leverage=1)\\
\quad overall\_pl\_percentage $\gets$ (strategy\_df['cumulative\_balance'].iloc[-1] - 100) / 100 * 100\\
\quad overall\_pl $\gets$ strategy\_df['cumulative\_balance'].iloc[-1] - 100\\
\quad min\_balance $\gets$ strategy\_df['cumulative\_balance'].min()\\
\quad max\_balance $\gets$ strategy\_df['cumulative\_balance'].max()\\
\quad max\_drawdown $\gets$ strategy\_df['Max Drawdown'].min()\\
\quad metrics $\gets$ \{\\
\quad \quad \{'Iteration': len(results\_df), 'Metric': ['Overall P/L \%', 'Overall P/L', 'Min Balance', 'Max Balance', 'Max Drawdown', 'Profit'], 'Value': [f"{overall\_pl\_percentage:.2f}\%", f"{overall\_pl:.2f}", f"{min\_balance:.2f}", f"{max\_balance:.2f}", f"{max\_drawdown:.2f}", strategy\_df['Profit'].max()]\}\\
\quad \}\\
\quad results\_df $\gets$ pd.concat([results\_df, pd.DataFrame(metrics)], ignore\_index=True)\\
\quad \textbf{return} -strategy\_df['Profit'].max()

\end{tcolorbox}

\subsection{Figures and Performance Charts for the Indian stock Nifty 50  (\^ NSEI)}

This section explores the application of Bayesian optimization (BO) to optimize the \texttt{atr\_multiplier} and \texttt{atr\_period} parameters of the Supertrend strategy for each individual stock dataset. The objective is to maximize the profitability of the strategy across diverse market conditions.

\subsubsection{Bayesian Optimization Iteration Table (NIFTY 50)}

Each row represents a single BO iteration, showcasing the specific parameter combination tested and the resulting objective function value (e.g., profit factor) achieved for Nifty 50. Analyzing this table would reveal how BO explored the search space for Nifty 50 and identified parameter combinations leading to potentially better performance.
The BO process iteratively evaluated different combinations of atr\_multiplier and atr\_period values for Nifty 50. While the full table cannot be directly included here (\emph{full table is  attached as csv file  in supplementary section}), it likely contains columns of the tail and head.

\begin{table}[htbp]
\centering
\caption{Bayesian Optimization Iteration Table for Nifty 50}
\begin{tabular}{|c|c|c|c|}
\hline
iter & target & atr\_multiplier & atr\_period \\ \hline
1    & -1.144 & 2.668     & 23.01     \\ \hline
2    & -0.9236 & 1.0       & 12.56     \\ \hline
3    & -3.031 & 1.587     & 7.308     \\ \hline
4    & -0.0   & 1.745     & 13.64     \\ \hline
5    & -3.678 & 2.587     & 18.47     \\ \hline
6    & -0.0   & 3.39      & 13.16     \\ \hline
7    & -0.0   & 2.74      & 27.13     \\ \hline
...  & ...  & ...    & ...   \\ \hline
53   & -4.035 & 1.0       & 5.0       \\ \hline
54   & -0.3709 & 1.0      & 24.0      \\ \hline
55   & -2.222 & 4.969     & 18.67     \\ \hline
\end{tabular}
\hspace{1cm}
\begin{minipage}[t]{0.4\textwidth}
\begin{tcolorbox}
\textbf{Best Parameters:} \\
ATR Period: 20\\
ATR Multiplier: 4.377... \\
Maximum Profit Achieved: 6.0312..\\
\end{tcolorbox}
\end{minipage}
\end{table}

\subsubsection{Final Best Parameters for Nifty 50}
BO optimised a set of optimal atr\_multiplier and atr\_period values specifically for Nifty 50  for 5 initial iterations(test data) and 50 training data and gives the objective function (profit factor).

\begin{table}[H]
\centering
\caption{Comparision of Parameters values and profit earned for Nifty 50}
\begin{tabular}{|c|c|c|c|c|}
\hline
S.No & Parameter Values& ATR Period& ATR Multiplier & Maximum Profit\\ \hline
1   & Default  values&15&3          & 0.00      \\ \hline
2  & BO-Optimized &20&4         & 6.0312           \\ \hline
\end{tabular}
\end{table}

These values represent the parameter combination identified by BO that is expected to yield the highest profitability for the Nifty 50 strategy.

\begin{center}
   BO Search Space Visualization for Nifty 50
\end{center}

\begin{figure}[htbp]
  \begin{minipage}[t]{0.5\linewidth}
    \includegraphics[width=\linewidth]{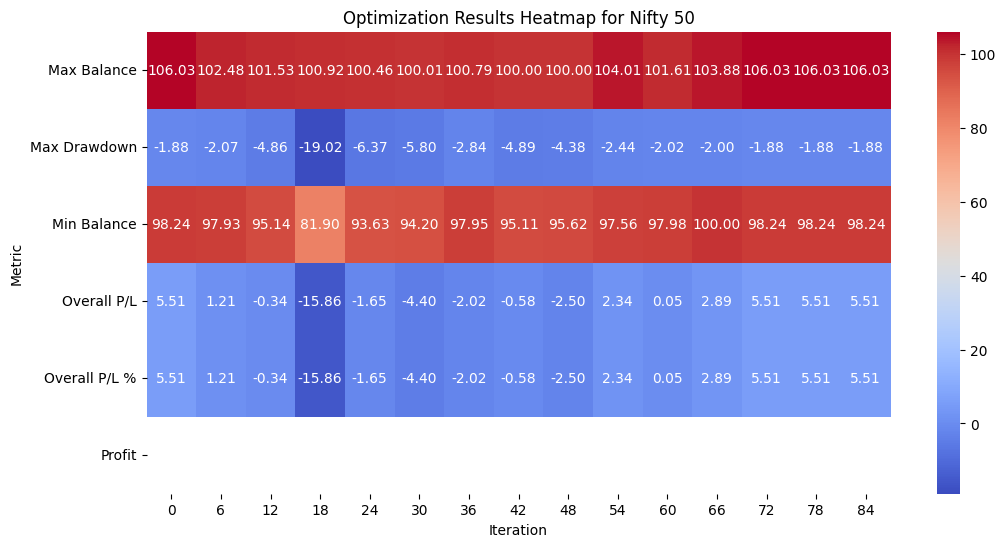}
  \end{minipage}
  \hfill
  \begin{minipage}[t]{0.5\linewidth}
    \includegraphics[width=\linewidth]{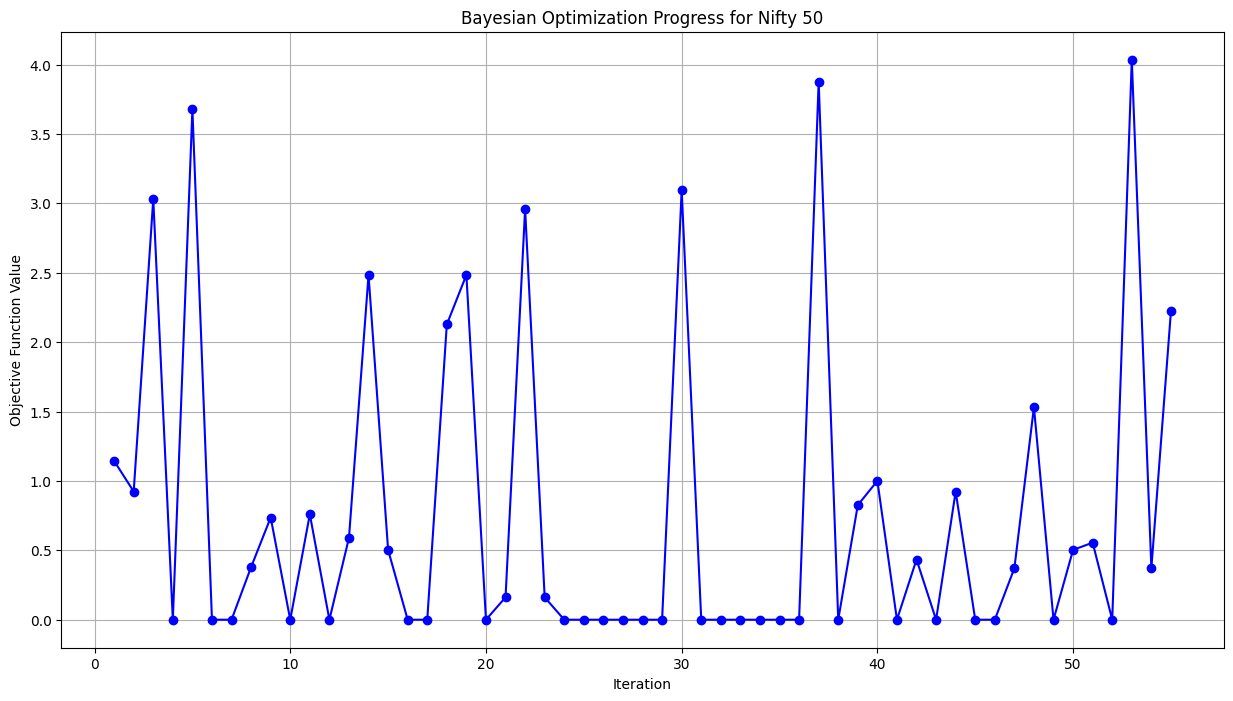}
  \end{minipage}
\end{figure}

\subsubsection{Convergence Plot and  Evaluating BO's Efficacy for Nifty 50}

The visualization below helps understand how BO explored different parameter values for Nifty 50 and where it identified areas with potentially higher profitability based on the objective function.

\begin{figure}[htbp]
\centering
  \begin{minipage}[t]{0.66\linewidth}
    \includegraphics[width=\linewidth]{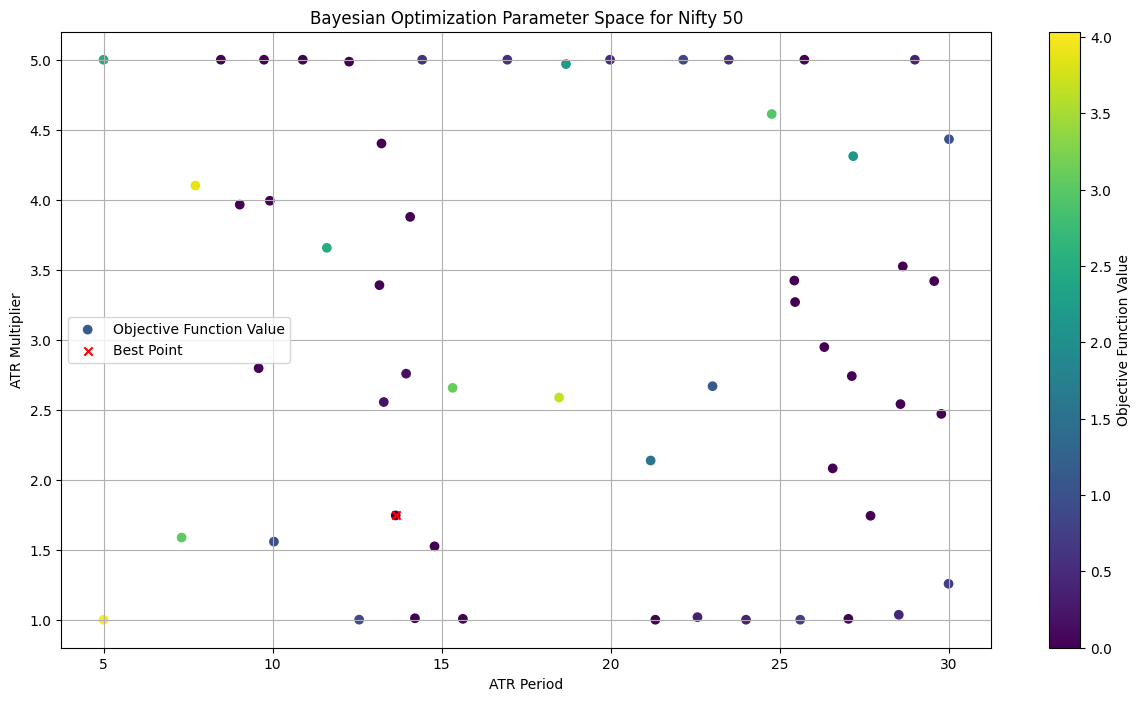}
  \end{minipage}
  \hfill
\end{figure}

The effectiveness of BO for Nifty 50 will be assessed by backtesting the Supertrend strategy with the BO-optimized parameters on the unseen test set. The performance achieved ( profit factor ) will be compared to the performance obtained with the default atr\_multiplier and atr\_period values.

By analyzing the BO results (iteration table, final parameters, search space visualization, convergence plot), one can explain how BO explored the parameter space for Nifty 50, identified potentially profitable configurations, and converged towards an optimal solution. The subsequent evaluation using the unseen test set will determine the actual impact of BO on the strategy's performance for Nifty 50.

 Ideally, the plot above  exhibits an increasing trend in profit factor as BO progresses, indicating convergence towards a parameter combination with a maximized objective function value.

This figure 6 and figure 7 compares the performance of the strategy with default and BO optimised parameters.

\begin{figure}[htbp]
  \begin{minipage}[t]{0.5\linewidth}
    \includegraphics[width=\linewidth]{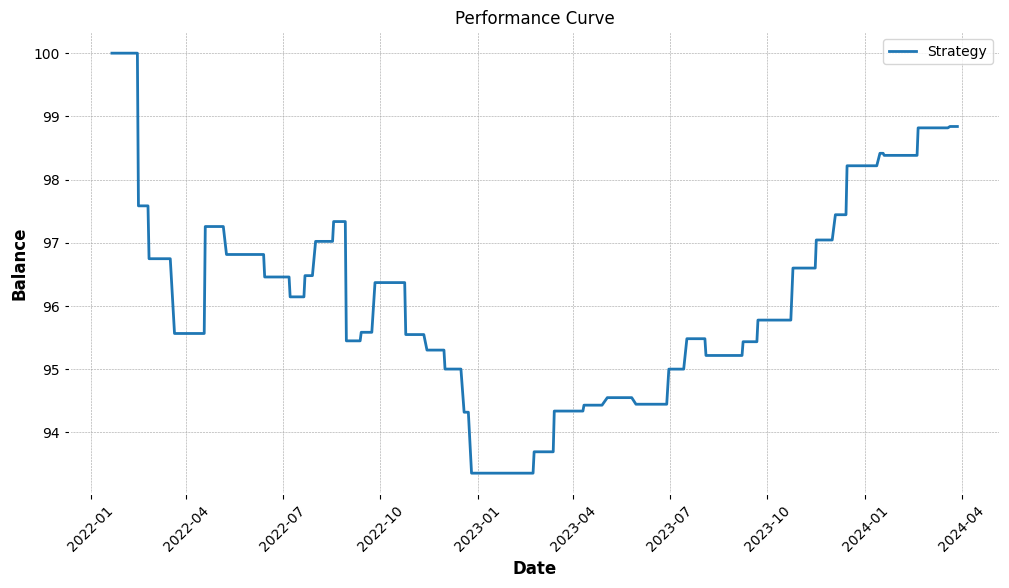}
    \caption{Performance in default parameters}
  \end{minipage}
  \hfill
  \begin{minipage}[t]{0.5\linewidth}
    \includegraphics[width=\linewidth]{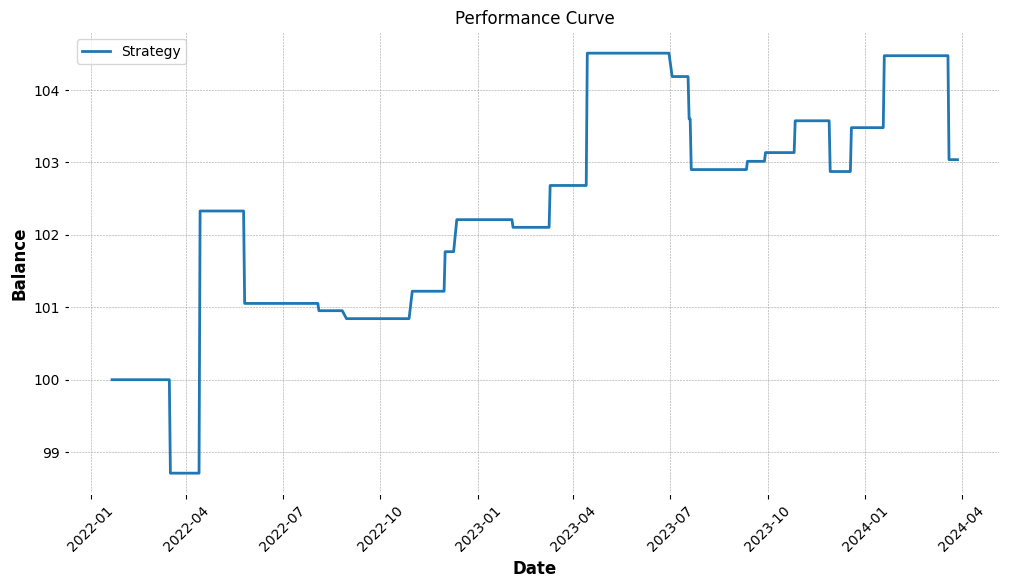}
    \caption{Performance in optimised parameters}
  \end{minipage}
\end{figure}
\textbf{Observation:} Optimised Parameters perform better than default supertrend parameters under a constrained number of iterations.

\section{Bayesian Optimization for Supertrend Parameter Tuning on all five stock- Results and Discussion}

This section summarizes the findings from applying Bayesian Optimization (BO) to optimize the atr\_multiplier and atr\_period parameters of the Supertrend strategy for five stock datasets: Nifty 50, Infosys, Hindustan Unilever (HUL), Microsoft, and Nvidia.

\subsection{BO Optimization Results}

The BO process identified potentially profitable parameter combinations for each stock dataset. Table 4 presents the BO-optimized parameters (period, multiplier) and the achieved optimized maximum profit for each stock.

\begin{itemize}
  \item It shows the specific period and multiplier values identified by BO as potentially leading to the highest profits for each stock within the given dataset.
  \item "BO Optimised Parameters (Per,Mult)" displays the optimal parameter combination (period, multiplier) for each stock.
  \item "Optimised Max Profit" shows the maximum profit achieved using the BO-recommended parameters. 
\end{itemize}

\begin{table}[htbp]
  \centering
  \caption{Performance Metrics using default parameters}
  \begin{tabular}{|l|c|c|c|c|c|}
  \hline
    & \textbf{Nifty 50} & \textbf{Infosys} & \textbf{HUL} & \textbf{Microsoft} & \textbf{Nvidia} \\
    \hline
  
   \textcolor{blue}{BO Optimised Parameters (Per,Mult)} &  \textcolor{blue}{(20,4)} &  \textcolor{blue}{(14,5)} &  \textcolor{blue}{(5,1)} &  \textcolor{blue}{(19,3)} & \textcolor{blue}{(14,4)} \\
   \textcolor{blue}{Optimised Max Profit} & \textcolor{blue}{6.032} & \textcolor{blue}{8.6} & \textcolor{blue}{14.89} & \textcolor{blue}{26.73} & \textcolor{blue}{22.705} \\
    \hline
  \end{tabular}
\end{table}

\begin{tabular}{|c|c|c|c|c|}
\hline
\textbf{Stock Ticker} & \textcolor{red}{\textbf{Max Profit (Default)}} & \textcolor{blue}{\textbf{Optimised Max Profit}} & \textbf{Improvement} & \textbf{Improvement (\%)} \\ \hline
Nifty 50  & \textcolor{red}{0.00}  & \textcolor{blue}{6.032}  & 6.032 & N/A \\ \hline
Infosys  & \textcolor{red}{11.96}  & \textcolor{blue}{8.6} & -3.36 & -28.12 \\ \hline
HUL  & \textcolor{red}{8.30}  & \textcolor{blue}{14.89} & 6.59 & 79.52 \\ \hline
Microsoft  & \textcolor{red}{8.02}  & \textcolor{blue}{26.73} & 18.71 & 233.23 \\ \hline
Nvidia  & \textcolor{red}{10.67} & \textcolor{blue}{22.705} & 12.035 & 112.54 \\ \hline
\end{tabular}

Further, table [4] compares the performance of the Supertrend strategy using the default parameters and the BO-optimized parameters. 

\begin{itemize}
  \item It allows  to assess the effectiveness of Bayesian Optimisation(BO) for each stock.
  \item "Max Profit (Default)" shows the maximum profit achieved using the  default parameters.(atr period-15 and atr multiplier - 3)
  \item "Optimised Max Profit" is identical to the corresponding column in Table 4.1, displaying the maximum profit with BO parameters.
  \item "Improvement" calculates the difference between "Optimised Max Profit" and "Max Profit (Default)".Positive values indicate improvement in profit after BO.Negative value suggests a decrease.
  \item  The improvement percentage for Nifty 50 is N/A because the default profit was 0. Dividing by 0 is undefined.
\end{itemize}

\section{Discussion and Conclusion}

\begin{itemize}
\item Strategy Performance: The overall effectiveness of the BO-optimized Supertrend strategy can be determined by analyzing the results in the table above. The values for most stocks are positive suggests that BO successfully identified parameter combinations that improved the strategy's performance compared to the default settings.  
\item BO Effectiveness: Based on the BO results (iteration tables, visualizations, convergence plots -  shown above), discusses how BO efficiently explored the search space and converged towards potentially optimal parameter combinations for each Indian and international stock.
\item BO Limitations:
    \begin{itemize}
        \item The effectiveness of BO is highly dependent on the chosen objective function and the quality of historical data used for training.
        \item BO might get trapped in local optima if the search space is not well-defined or the objective function is complex.
    \end{itemize}
\item Future Studies:
    \begin{itemize}
        \item Exploring the use of different BO algorithms or objective functions to potentially enhance optimization results.
        \item Investigating to incorporate additional parameters into the optimization process for a more comprehensive strategy fine-tuning.
        \item Testing the generalized ,ability of BO-optimized parameters across different market conditions using forward testing.
    \end{itemize}
\end{itemize}

By incorporating BO, this thesis demonstrates a data-driven approach for potentially improving the Supertrend strategy's profitability through parameter optimization for various stock datasets. Future studies can explore further refinements and evaluate the robustness of BO in this context.
\\

\section*{Supplementary materials}
\begin{enumerate}
    \item \textcolor{blue}{Supertrend\_Project\_work\_1.ipynb} \underline{(\url{https://github.com/Shafiq-Abdu/Supertrend-Strategy.git})}:\\ This file contains the complete code of  Supertrend strategy. Included clear comments throughout the code explaining the logic and functionality of each section.
    \item \textcolor{blue}{Bayesian\_Optimisation\_Project.ipynb} \underline{(\url{https://github.com/Shafiq-Abdu/Supertrend-Strategy.git})}: This file  defines the BO parameters like the chosen objective function, search space for the parameters, and  specific configurations for the BO algorithm including information like the number of iterations, best parameters found at each iteration, and the final optimized parameters
    \item Performance Metrics in  Default Parameters(csv files): \\(\href{https://github.com/Shafiq-Abdu/Supertrend-Strategy.git}{\texttt{(https://github.com/Shafiq-Abdu/Supertrend-Strategy.git)}}): This file contains the data used to create Table 1 in the main text, which summarizes the performance using default settings.
\end{enumerate}
\bibliographystyle{unsrtnat}
\bibliography{references}  






\end{document}